\documentclass[10pt, preprint]{aastex}
\usepackage{natbib,color,lscape}
\title{Population boundaries for compact white-dwarf binaries in
LISA's amplitude-frequency domain}
\author{Ravi kumar Kopparapu\altaffilmark{1}}
\altaffiltext{1}{Center for Gravitational Wave Physics, 104 Davey lab, Pennsylvania State
University, University Park, PA - 16802-6300, USA}
\begin{document}
\begin{abstract}
In an earlier investigation, we proposed population boundaries for 
both inspiralling and mass-transferring double
white dwarf (DWD) systems in the distance independent ``absolute'' amplitude-frequency domain
of the proposed space-based gravitational-wave (GW) detector, {\it LISA}.
 The degenerate zero temperature mass-radius (M-R) relationship of individual white dwarf stars that
 we assumed, in combination with the constraints imposed by Roche geometries, permits us to identify
 five key population boundaries for DWD systems in various phases of evolution.
Here we use the  non-zero entropy donor M-R relations of \cite{DB2003} to modify these boundaries for 
 both DWD and neutron star-white dwarf (NSWD) binary systems. We find that the mass-transferring
systems occupy a larger fraction of space in ``absolute'' amplitude-frequency domain compared to
the simpler $T=0$ donor model. We also discuss how these boundaries are
modified with the  new evolutionary phases found by \cite{Deloyeetal2007}. In the initial
contact phase, we find that the
 contact boundaries, which are the result of end of inspiral evolution,  would have some width, as opposed to an abrupt cut-off described in our earlier 
$T=0$ model.
 This will cause an overlap between a
DWDs $\&$ NSWDs evolutionary trajectories, making them indistinguishable with 
only LISA observations
within this region.  In the cooling phase of
the donor, which follows after the adiabatic donor evolution, the radius contracts,
mass-transfer rate drops and slows down the
orbital period evolution. Depending upon the entropy of the donor,
 these systems may then lie
inside the fully degenerate $T=0$ boundaries, but LISA may be unable to
 detect these systems as they might
 be below
the sensitivity limit or within the unresolved DWD background noise.
 We assess the limits and applicability of our theoretical population boundaries
with respect to observations and find that a measurement of $\dot f$ by LISA at high frequencies 
(Log $[f] \geq 2$) would likely distinguish between DWD/NSWD binary. For low frequency
sources, GW observations
alone would unlikely tell us about the binary components, without the help of electromagnetic
 observations. 
\end{abstract}

\section{Introduction}
\label{sec1}
The proposed space-based Gravitational-wave (GW) detector, 
{\it LISA}\footnote{http://lisa.nasa.gov} ({\it Laser Interferometer Space Antenna})
\citep{FB84,EIS87,Bender98}, is sensitive to GWs in the $10^{-4} - 1$ Hz frequency range. 
Within this band, one of the most promising sources are double white dwarf (DWD)
 binary systems, as it is expected that a large fraction of
 main-sequence binaries 
end their lives as close DWDs \citep{IT84, IT86}. For this reason, the GWs emitted by these systems in our Galaxy 
may form a background noise in
the low frequency ($\leq 3 \times 10^{-3}$ Hz) band of {\it LISA}. The population of DWDs in 
our Galaxy is 
expected to be dominated by systems that undergo two distinct,
long-lived phases of evolution: an ``inspiral'' phase, where both the stars are detached from
their Roche lobes and the loss of angular momentum in the form of GW emission
 causes the two stars to slowly spiral in towards each other; and
 a ``stable mass transfer'' phase, where the less massive star fills its
Roche lobe initially and starts transferring mass steadily to its companion. An example of 
the stable mass transferring systems are the AM CVn type systems, of which 18
 \citep{Nelemans2005} are known through electromagnetic observations\footnote{Two
controversial candidate systems, RX J0806+15 and V407 Vul may change their number to $16$. See
\cite{Cropper1998, Wu2002, MS2002} for more details.}.

Apart from DWDs, neutron star white dwarf (NSWD)
 binary systems are also  one of the promising sources of GWs for LISA.
Various authors \citep{Kim2004, cooray2004, Nelemans2001} have estimated the GW background from
 these systems and concluded
that the number of NSWD systems detectable with LISA is 1-2 orders of magnitude less than DWD systems.
 Similar to DWDs, NSWD systems also undergo 
inspiral and stable mass transfer phases. Specifically, several studies \citep[see for example:][]{Nelson1986,
 BT2004a, Nelemans2006}
 have suggested that one of the possible formation scenarios of the so-called 
ultra compact x-ray binary (UCXB) systems,
with orbital periods  $\le 80$ minutes, is that a low mass
white-dwarf donor ($\le 0.1 M_\mathrm{\odot}$) transferring mass to an accreting 
neutron star (NS) primary in a short orbit.
In this scenario, a detached NSWD binary system initially evolves to a minimum orbital period as  angular
momentum is lost from the system due to GW radiation. At this minimum orbital period, the companion white dwarf (WD) star
starts filling its Roche lobe and transfers mass to the NS  and the system evolves to longer orbital periods.
At present, there are $12$ known UCXB systems with measured orbital periods, 
see Table\ref{table1}. Some of these systems have accreting millisecond pulsars
(XTE J1807-294, XTE J1751-305 \& XTE J0929-314: see \cite{Mark2002}, for example)
 and several of them are found in globular clusters as the stellar density and
 close encounters are more common \citep{clark1975,ivanova2007}. The total 
population of field UCXBs may be low, $\sim 10$ \citep{BT2004a, cooray2004} and 
theoretical studies by \cite{BT2004b} indicates that even at the Galactic center,
 accreting NS systems do not contribute much to the faint x-ray population.

In general, the capabilities of {\it LISA} as a GW detector are usually discussed in
the context of the log $(h)$-- log $(f)$ domain.
 \cite{KT2007} proposed  that an analogy can be drawn between the astronomy 
community's familiar color-magnitude (CM) diagram and {\it LISA}'s 
 amplitude-frequency  diagram.
 For LISA
sources, an analogous quantity to absolute magnitude $M$ is $\log(rh)$, where $r$ is the 
distance to the source.
The underlying
physical properties of compact binary  systems such as DWDs and NSWDs,
 their evolution, and their
relationship to one another in the context of stellar populations
can be ascertained only if the observational properties of such
systems are displayed in a $\log(rh) - \log(f)$ diagram, rather than
in a plot of $\log(h)$ versus $\log(f)$. \cite{KT2007} discussed the DWD
binary systems in this context of ``absolute'' amplitude-frequency domain assuming that
the donors in these systems follow zero temperature mass-radius relation. Here, 
we will consider the effect of ``warm'' donors as proposed
 in \cite{DB2003, Deloyeetal2005, 
Deloyeetal2007}
and extend the discussion to NSWD binary systems and compare the population
boundaries between both of them. We also discuss the limits and applicability of
these population boundaries in the log$(rh)$ - log$(f)$ space within the context of
 observed AM CVn and UCXB systems.

\section{Evolution of WD binaries in the amplitude-frequency domain}
\label{sec2}
As mentioned in the introduction,
since the donor is a WD star and starts filling its Roche lobe to begin  mass transfer phase, 
 we will denote the donor WD with subscript {\it d} and the
accreting companion  as {\it a}. This notation
will be followed even during the detached inspiral phase of the evolution.
 Hence, the total mass $M_\mathrm{tot} = M_\mathrm{d} + M_\mathrm{a}$ and the mass ratio $q = M_\mathrm{d} / 
M_\mathrm{a}$. 
 Also we assume that the maximum mass of a WD to be Chandrasekhar mass, $M_\mathrm{ch}=1.44 M_\odot$, the
minimum mass of a NS  is $1.2  M_\odot$ and the maximum mass of NS to be $3  M_\odot$.

In the case of a detached DWD or a NSWD system inspiralling as a result of loss of
angular momentum due to gravitational radiation, we can write,
 as shown in \cite{KT2007}, as
\begin{eqnarray}
rh_\mathrm{norm} &=& \biggl[ \frac{2^5\pi^2}{c^2} \biggl(
\frac{GM_\mathrm{ch}}{c^2} \biggr)^5  K^5 f^2 \biggr]^{1/3} = 5.38
~[K^5 f^2]^{1/3}~\mathrm{m} \, , \label{h_f_relationship}
\end{eqnarray}
where the dimensionless mass parameter,
\begin{eqnarray}\label{K_definition}
K \equiv 2^{1/5}\biggl( \frac{\mathcal{M}}{M_\mathrm{ch}} \biggr) =
2^{1/5} \biggl( \frac{M_\mathrm{tot}}{M_\mathrm{ch}}  \biggr)
Q^{3/5} = \biggl(\frac{M_\mathrm{a}}{M_\mathrm{ch}}\biggr) \biggl(
\frac{2 q^3}{1+q}\biggr)^{1/5} \, ,
\end{eqnarray}
$\mathcal{M} = M_\mathrm{tot}Q^{3/5}$ is the chirp mass and $Q \equiv q/(1+q)^2$.
Notice that, for DWDs,
 the maximum value of $K = 1$ occurs at $\mathcal{M} = 1.25 M_\odot$
 ($q=1, M_\mathrm{tot} = 2.88 M_\odot$)
but for NSWD 
binaries, the maximum value of $K$ is $1.39$ and occurs at $\mathcal{M} = 1.75 M_\odot$
($q=0.5, M_\mathrm{tot} = 4.32 M_\odot$). This is purely due to the upper limit on the 
mass of the WD, $M_\mathrm{ch}$. Similar to a DWD system, where a $ K =1$
 creates a maximum inspiral trajectory in the log$[rh_\mathrm{norm}]-$log$[f]$
plane,  a
 $K = 1.39$ makes the limiting inspiral trajectory for NSWD binary systems 
in this plane.
This limiting trajectory, for both DWD $\&$ NSWD is plotted in  Figs. \ref{fig1} $\&$ \ref{fig2},
respectively, as a red line with a 
slope of $2/3$ (see Eq.(\ref{h_f_relationship}))
 beyond which a detached inspiralling DWD or NSWD system can not be found.

The phase of a detached inspiral evolution terminates when the low mass WD companion comes
 into contact with it's
Roche lobe to initiate a phase of mass-transfer.  Since the
 donor is a WD star this contact period can be found by equating the radius of the
 star  with the Roche lobe radius.  Here, for the radius of the WD, we use 
 the mass-radius (M-R) relationship of non-zero entropy donor models of  \cite{DB2003}.
These models were initially developed
for ultra-compact x-ray binaries (UCXB) with WD donors.
Later \cite{Deloyeetal2005} applied them 
 to model the donors in AM CVn systems formed through WD channel.
For these models the donors are assumed to be fully convective and  corresponding
donor evolution is considered to be adiabatic in nature due to the large mass-
 transfer rates produced in AMCVn systems. However, \cite{Deloyeetal2007}
 showed that
the assumption of adiabatic evolution is applicable during the beginning of mass-transfer phases
 (when the mass-transfer rates are high) and may not work when the system evolves to larger 
orbital periods with low mass-transfer rates. Instead, they have identified new evolutionary
phases and we will also discuss the impact of this most recent study 
 on our population boundaries. We illustrate these boundaries by assuming that the donor
 composition is He, but it is straight forward to extend the discussion to
carbon (C) $\&$ oxygen (O) donors.
In this section, we will discuss in detail the population boundaries of both DWD 
$\&$ NSWD. In \S\ref{sec3} we will discuss the applicability of these boundaries
with respect to LISA observations, along  with the effects of new phases of evolution
 detailed in \cite{Deloyeetal2007}. Finally, we summarize in \S\ref{sec4}.

\subsection{Population boundaries for DWD systems}
\begin{figure}[!hbp|t]
\epsscale{0.8}
\includegraphics[height=13.0909cm,width=15.8779cm,angle=360]{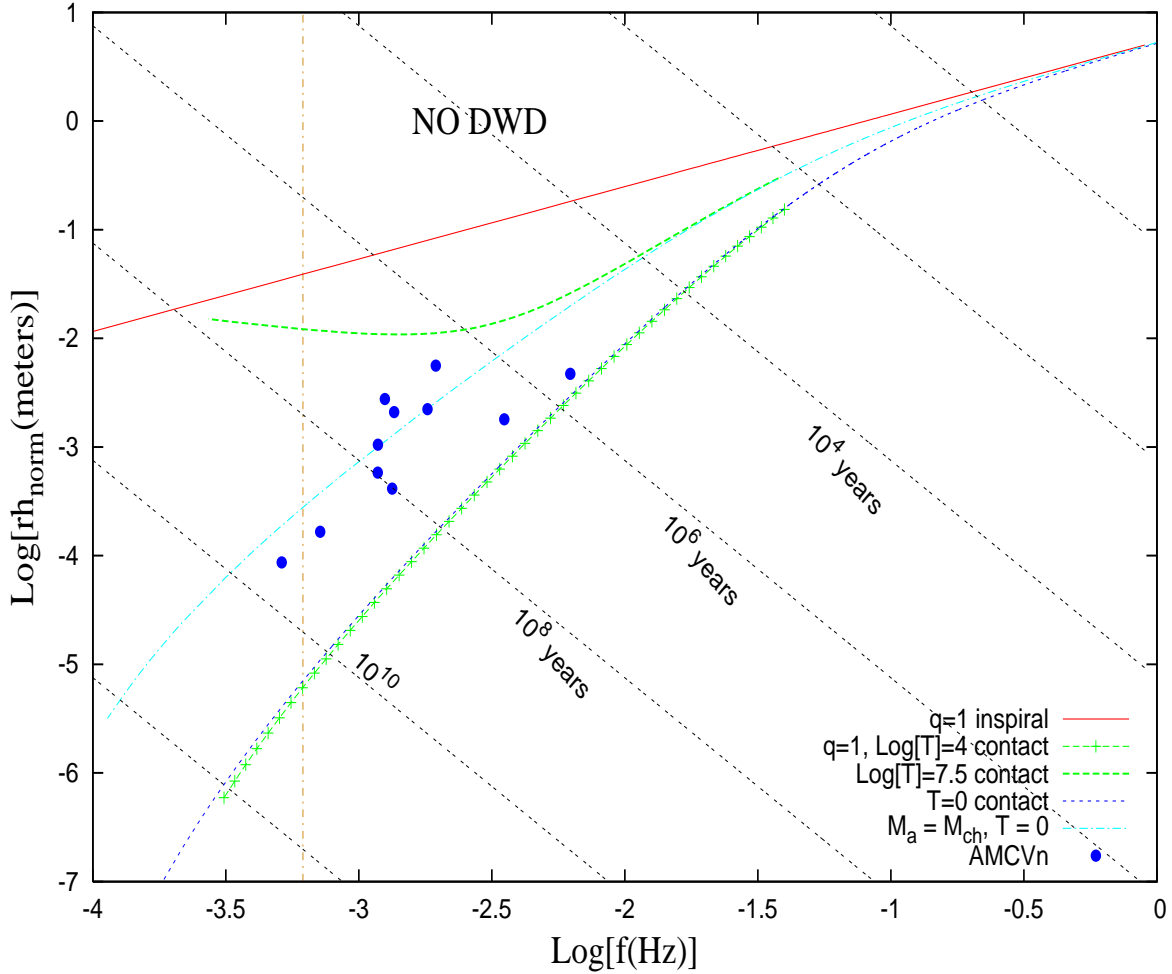}
\hfill \caption{ Population boundaries for DWD systems. The (red) line with positive
slope represents the maximum inspiral evolution trajectory for DWD systems, beyond which 
none will be found. This happens at $q = 1, M_\mathrm{tot} = 2.88 M_\odot$. The bottom
(green) curve with crosses is drawn using the M-R relation of isentropic donor models
from \cite{DB2003}. It represents the termination (contact) boundary for $q=1$ 
systems, assuming that the donors are of He composition with $T=10^{4} K$. For 
comparison, contact boundary derived using Eggleton's $T=0$ M-R relation for fully
 degenerate He donors is also shown as (blue) dashed line. The top (green) dashed 
curve represents the boundary at which $M_\mathrm{a} = M_\mathrm{ch}$, assuming that
 donors in DWD systems would have a  maximum $T = 10^{7.5} K$. No mass-transferring DWD
would be found above this boundary. For comparison, the (blue) dot-dashed curve shows 
a similar boundary for $T=0$ donor models. Also shown are constant $\tau_\mathrm{chirp}$
lines along with observed AMCVn systems (blue dots; see text). 
Explanation for vertical
(brown) dot-dashed line is given in \S\ref{sec3}.
}
\label{fig1}
\end{figure}

In Fig. {\ref{fig1}},  We show the population boundaries for both inspiralling $\&$ 
mass-transferring phases of DWD systems. The (red) line with slope $=2/3$ shows the
maximum inspiral trajectory ($q=1, M_\mathrm{tot} = 2.88 M_\odot$) from a detached DWD system.
Above this boundary, no DWD binary can exist.
The lower (green cross) curve shows the locus of
the termination (contact) points of all detached inspiralling DWD binary systems, which
have $q=1$, assuming that the WD is a He donor and has a temperature of $T = 10^{4} K$ (which is 
equivalent to the boundary obtained with $T=0$ M-R relation; along this
isotherm, $M_\mathrm{tot}$ decreases from top to bottom). For 
comparison, we have also shown the contact boundary for $q=1$, generated by 
Eggleton's mass-radius (M-R) relationship for $T=0$ white dwarfs, (blue
dashed line), as quoted by \cite{VR88}
and also by \cite{MNS}.
 At higher amplitudes and frequencies, both the curves overlap each other.
This is due to the overlap of  M-R relations at higher masses as
these objects are supported by degeneracy pressure and the $T=0$ M-R relation that we
are using is applicable for fully degenerate He donors.
At lower amplitude and frequencies, the contact boundaries diverge slightly.
This is because there is a turn over in the M-R relation at low masses for $T=10^{4}K$ as 
the Coulomb interactions
dominate thermal contribution at that low temperatures and the  M-R relations deviate.
So this ($q=1$) contact curve represents the boundary beyond which no DWDs with He 
donors will be found. The corresponding boundaries for C $\&$ O would lie just below
this contact boundary, except that all three (He, C $\&$ O) boundaries merge at 
high frequencies and deviate at low frequencies.

Once the mass transfer phase is initiated, the orbital separation $a$ starts increasing
 and the system evolves to lower
amplitudes and frequencies. The evolutionary trajectories can be then traced out
 by assuming that
the donor WD star is marginally in contact with it's Roche lobe and conservative mass 
transfer (CMT) holds true.
 For DWDs, Since we assume that the maximum mass in a DWD system
 is $M_\mathrm{ch}$,   we can
 generate a upper boundary beyond which no mass-transferring DWD system can exist.
This limiting boundary for DWDs can be
obtained for systems with $M_\mathrm{tot} > M_\mathrm{ch}$ (because, then,
 eventually $M_\mathrm{a}$
will exceed $M_\mathrm{ch}$) when their $q$ drops below a critical $q_\mathrm{ch}$
\begin{eqnarray}
q_\mathrm{ch} = \frac{M_\mathrm{tot}}{M_\mathrm{ch}} - 1
\label{qch}
\end{eqnarray}
 Accordingly, the (green) dashed line 
in Fig. \ref{fig1} indicates an isotherm with Log$[T]=7.5$ and
 $M_\mathrm{a} = 1.44 M_\odot$. We note that Log$[T]=7.5$ 
 upper boundary is likely a
generous overestimate of the available phase space even if the donors were
to evolve adiabatically from contact.
Assuming that there are no donor WDs in a DWD system with Log$[T] > 7.5$, this curve
 represents the
boundary beyond which no mass transferring DWDs can exist.  Since the \cite{DB2003} 
donor models have two branch nature, with a
minimum attainable mass, it is  reflected in this boundary as it curves up,
 with a minimum attainable GW amplitude for a given isotherm. 
Corresponding boundary for donors with C $\&$ O composition would lie below this
 He curve because a higher temperature is
required for a C/O donor to fill the Roche lobe than for a  He donor.
  For 
comparison, a similar boundary that arises from $T=0$ M-R relation is also shown as
(light blue) dot-dashed curve.
 As is the case with $q=1$ contact boundary, these two
curves overlap at high frequencies.

From the above discussion, we note that all mass-transferring DWDs with different
donor masses, composition and temperatures  should be constrained within these two
 boundaries\footnote{For C $\&$ O composition donors, as discussed above, there is a
slight deviation at the lower contact boundary.}.
 Compared to T=0 model, as assumed in \cite{KT2007}, hot donors in mass-transferring
DWDs occupy larger fraction of space in $rh-f$ domain. This is because
hot donors are also more massive, therefore have higher intrinsic GW amplitude and come into
contact at lower frequencies because of larger radius.

The inspiral phase of a WD binary evolution is driven by the loss of the 
angular momentum due to GW radiation and the inspiral evolutionary time scale 
$\tau_\mathrm{chirp}$ can be written as:
\begin{eqnarray}
\tau_\mathrm{chirp} &\equiv& \frac{5}{256} \frac{c^5 a^4}{G^3
M_\mathrm{tot}^{3}} \biggl[\frac{(1+q)^2}{q}\biggr] = \frac{ 5}{64
\pi^2 } \biggl( \frac{c}{rh_\mathrm{norm} f^2} \biggr) \, . \label{tau_chirp}
\end{eqnarray}
It is worth noting that the time scale of evolution in both inspiral and 
CMT phases is of the order of $\sim \tau_\mathrm{chirp}$ \citep{KT2007}. 
Therefore we have drawn ``chirp'' isochrones also in Fig.\ref{fig1} and each
isochrone has a slope of $-2$ in this log$(rh_\mathrm{norm})$ - log$(f)$ space,
as can be seen from Eq.(\ref{tau_chirp}) due to the dependence on the product
of $rh_\mathrm{norm}$ and $f^{2}$. This implies that a given WD binary system, 
in either inspiral or CMT phase, spends $\tau_\mathrm{chirp}$ amount of time
 corresponding to the GW frequency that the binary is emitting.
 At lower frequencies, the value of $\tau_\mathrm{chirp}$ is larger compared to
that of at higher frequencies. A consequence of this behavior is that more
 binary systems accumulate at lower frequencies. As mentioned in the
 introduction, the background noise 
arising due to millions of DWDs at the lower frequency band ($\leq 3 \times 
10^{-3}$ Hz) of LISA is due to the fact that the value of $\tau_\mathrm{chirp}$
for these systems is $\sim 10^{10}$ years, approaching Hubble time (see 
Fig.\ref{fig2}).

 
\clearpage
\thispagestyle{empty}
\clearpage

\subsection{Population boundaries for NSWD systems}
\begin{figure}[!hbp|t]
\includegraphics[height=13.0909cm,width=15.8779cm,angle=360]{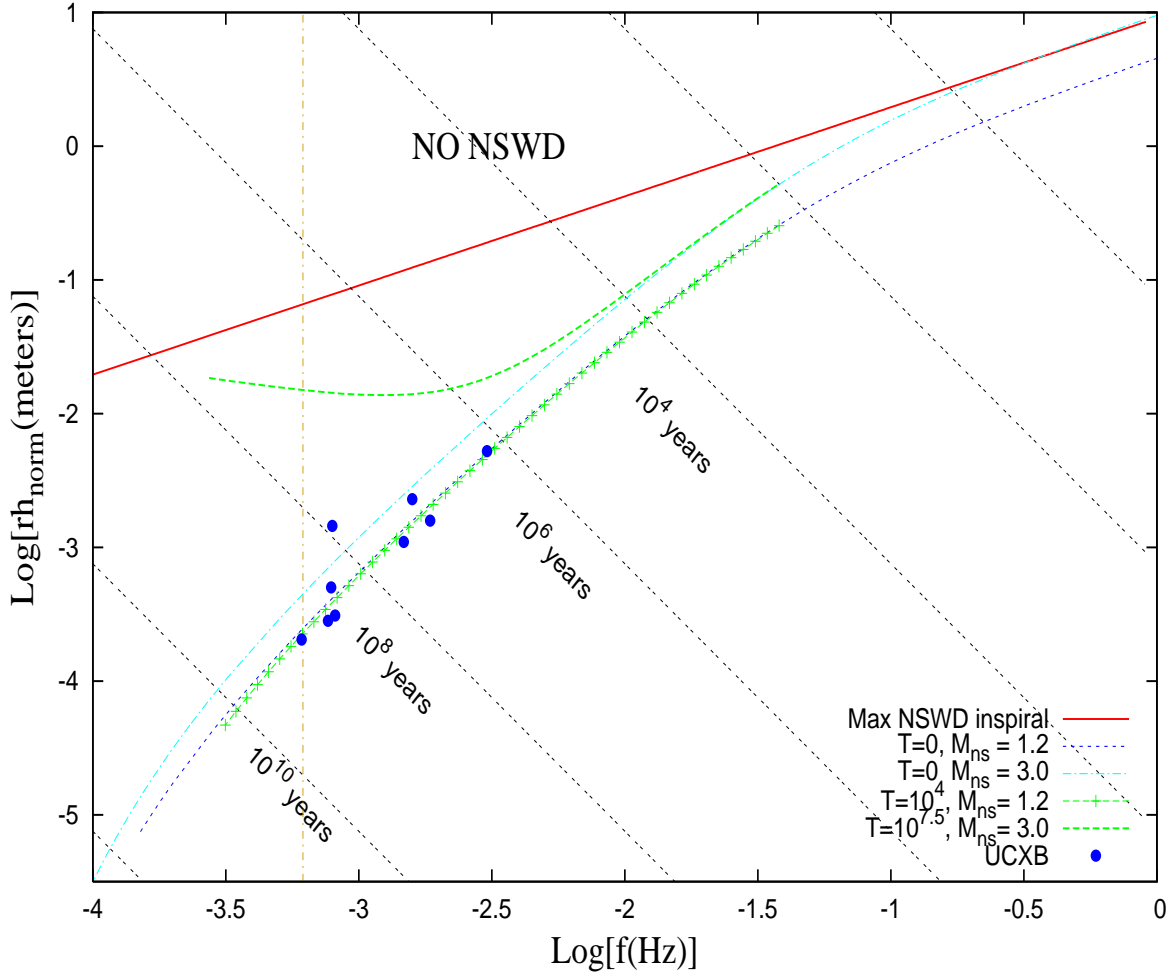}
\hfill \caption{ Population boundaries for NSWDs in the
 log$[rh_\mathrm{norm}]$-log$[f]$ plane.
The red line represents the maximum inspiral trajectory from a 
detached NSWD binary system ($q=0.5, M_\mathrm{mtot} = 4.32 M_\odot$). The bottom
curve with (green) crosses represents the contact boundary, assuming minimum mass of
the NS to be $1.2 M_\odot$ with  He donors of $T=10^{4} K$. The blue dashed line is
drawn assuming a $T=0$ M-R relation for fully degenerate He donors. These two boundaries
overlap at high masses (high amplitude $\&$ frequency) and deviate slightly at lower 
masses. The top (green) dashed curve shows the upper bound for mass-transferring NSWD
systems, assuming that the maximum donor temperature in a NSWD system to be $T=10^{7.5} K$
and maximum mass of NS as $3.0 ~ M_\odot$. To compare, an upper boundary drawn
using $T=0$ donor model is also shown as (light blue) dot-dashed curve.  Also shown are
currently known UCXB systems (blue dots) along with chirp isochrones.}
\label{fig2}
\end{figure}

In Fig. \ref{fig2}, population boundaries for both NSWD systems is shown in the 
log$(rh_\mathrm{norm})$ - log$(f)$ space. The top (red) curve represents the maximum
inspiral boundary for detached NSWD binary systems ($K = 1.39$) beyond which none will
be found. The bottom (green) curve with crosses represents the lower contact boundary,
drawn using \cite{DB2003} hot donor models, assuming that the donor is of He composition
 with $T=10^{4} K$. The mass of the 
NS on this boundary is set to the minimum mass assumed, $1.2 M_\odot$. For comparison,
a similar boundary using $T=0$ M-R relation for He donors
is also shown as (blue) dashed curve. As is the case with DWDs, these two boundaries
overlap at higher amplitudes and frequencies because at higher donor masses, the two
M-R relations match as they are supported by degeneracy pressure. Furthermore, there is an
overlap region between the lower NSWD contact boundary (green dashed curve in Fig. \ref{fig2})
 $\&$ the $M_\mathrm{a} = M_\mathrm{ch}$ DWD contact boundary
(green dashed curve in Fig. \ref{fig1}), indicating both types of systems can exist within
this region.  Similarly, an upper 
contact boundary can also be drawn for NSWDs, shown as (green) dashed curve in Fig. \ref{fig2}, 
assuming the maximum mass of the NS is $3.0 M_\odot$ and that the maximum donor 
 temperature in a NSWD system to be $T=10^{7.5} K$.
 The light blue dot-dashed curve is the upper contact boundary for $T=0$ donors.
Just like DWDs, for NSWD systems with $M_\mathrm{tot} > 3 M_\odot$, the critical
 mass ratio $q_\mathrm{NS}$ below which a NSWD binary can not exist, is
\begin{eqnarray}
q_\mathrm{NS} = \frac{M_\mathrm{tot}}{3 M_\odot} - 1
\label{qNS} 
\end{eqnarray}
So all the mass-transferring NSWD binary systems are bounded by the upper $\&$ lower
 (green) curves\footnote{The lower boundary is applicable for He donors.}
 and all the detached inspiralling NSWDs will be bounded between the 
top (red) curve and the bottom (green) curve.
Fig. \ref{fig2} also shows chirp isochrones,  indicating the evolutionary timescales.


\clearpage
\thispagestyle{empty}
\clearpage

\section{Discussion}
\label{sec3}
The population boundaries for DWD $\&$ NSWD binary systems discussed in
 previous sections were generated using \cite{DB2003} non-zero entropy donor
models,  which assume that the donors are fully convective and undergo
adiabatic evolution throughout  the mass-loss phase. Recently, \cite{Deloyeetal2007}
showed that these assumptions may not properly estimate the donor's orbital
period evolution and specifically the donor's adiabatic evolution may not hold true
 for whole mass-transfer phase. Instead they identified three distinct 
evolutionary phases and here we discuss the implications of their new findings
on our population boundaries.

During the first phase, which happens during the mass-transfer ``turn-on'' phase
(when the donor comes into contact initially), the radius of the donor decreases
, the mass-transfer rate $\dot M_\mathrm{d}$ increases and 
the orbital period $P_\mathrm{orb}$ continues to decrease until the donor 
radius reaches it's minimum value ($\dot M_\mathrm{d}$ becomes maximum)
and starts expanding again.
 \cite{Deloyeetal2007} calculated that this turn on phase
lasts up to $\sim  10^{6}$ years. 
In the second phase, the donor responds to the mass loss
adiabatically and starts expanding, which is considered to be the normal 
AM CVn phase. 
But this phase of adiabatic expansion ends and a third phase of evolution 
begins at around $P_\mathrm{orb} \sim 45$ min,
when the mass-transfer rate (and the donors thermal time) drop enough for
 the donor to cool and start contracting
to a fully degenerate configuration, stalling the $P_\mathrm{orb}$ evolution. 

In the case of DWDs, the initial turn-on phase will result in the $q=1$ contact
 boundary (green curve with crosses in Fig. \ref{fig1}) to 
 have a width, instead of a sharp boundary as discussed in \cite{KT2007}.
 This is because it is calculated assuming that once the system comes into
contact, it would evolve towards lower amplitudes and frequencies, whereas
 $P_\mathrm{orb}$ decreases in the turn on phase, to a minimum even after the
initial contact. Accordingly, there will be a slight overlap of lower contact boundaries
for He, C $\&$ O donors. Once the system evolves off this initial contact phase, the donor
expands adiabatically in response to the mass-transfer and the system follows a
typical AM CVn evolution,  where the GW amplitude $\&$ frequency keeps decreasing
as $P_\mathrm{orb}$ increases. Assuming an adiabatic evolution means that the cooling
time of the donor is longer than the mass-transfer time-scale 
($M_\mathrm{d} / \dot M_\mathrm{d}$), it will in turn affects the
orbital evolution time-scale. Accordingly, the donor follows a trajectory where it 
passes through different isotherms of decreasing $q$, after the initial contact. As \cite{KT2007} illustrate,
the lower right curve in Fig. \ref{fig1} shows the contact boundary for $q=1$ 
systems and similar contact boundaries for lower $q$'s would 
lie to the left of it, but the 
contact boundaries will shift towards lower amplitudes and frequencies.  
This phase of adiabatic evolution comes to an end between
 $P_\mathrm{orb} \approx 40-55$ min, when the donor starts to cool and contract 
eventually towards a fully degenerate  star. 
This will drastically (almost an order of magnitude, see \cite{Deloyeetal2007} Fig. 15)
reduce $\dot M_\mathrm{d}$ at these long orbital periods.
 Hence, after this range of $P_\mathrm{orb}$, the systems
GW frequency evolution slows down in accordance with the drop in $P_\mathrm{orb}$ evolution. If
the donor has cooled enough to approximate it as a $T=0$ degenerate model, then 
it may lie close to
one of the $T=0$ contact boundaries, depending upon its $q$. But these $T=0$
boundaries are bounded within the
region constrained by $M_\mathrm{a} = M_\mathrm{ch}$ (light blue dot-dashed) curve 
 and the $q=1$ contact boundary (green curve with crosses in Fig. \ref{fig1}), because these boundaries are drawn assuming
that the donor is a fully degenerate star with $T=0$ and such a system can not exist beyond
these curves. Therefore, we have drawn a vertical
(brown) dot-dashed line in Fig. \ref{fig1} at $P_\mathrm{orb} = 55$ min (Log$f = 3.21$) beyond which we would expect
these systems to be within the $T=0$ contact boundaries. Note that this vertical line
is not a sharp boundary: some systems may not reach a fully degenerate
 configuration by this $P_\mathrm{orb}$. Rather, a system's $P_\mathrm{orb}$
evolution slows down at a particular GW frequency
once they start cooling towards a degenerate configuration at the above mentioned $P_\mathrm{orb}$.
Fig. \ref{fig1} also shows the observed AM CVn type systems, for which the masses and $P_\mathrm{orb}$
 are taken from \cite{Deloyeetal2005} $\&$ \cite{Roelofs2007}. Couple of them are clearly outside $T=0$ 
region; for some of them, the system's mass function limits the minimum donor
mass to a value above that of a Roche-filling $T=0$ donor at the same 
$P_\mathrm{orb}$.  It is very difficult to know their exact temperature and/or
composition purely from GW observations, as there is a good overlap of systems with these 
characteristics\footnote{Out of these systems, 
CE 315 (the system with lowest GW amplitude and frequency in Fig. \ref{fig1})
 has $P_\mathrm{orb} = 65$ min and lies to the
left of the $P_\mathrm{orb} = 55$ line and below
 $T=0, M_\mathrm{a} = M_\mathrm{ch}$ boundary. This may seem to imply that this system may be
 cooling off and trying to reach a $T=0$ degenerate configuration. But in Fig 1. of 
\cite{Deloyeetal2005}, they give a temperature range of Log $T = 4.0-6.5$ and \cite{Bildsten2006} noted
that this system may have a hot donor.  If that is the case, \cite{Bildsten2006} note that this donor
may have evolved with constant entropy that it was born with or may have been heated by either the disk
or the accretor. However, if irradiation is the cause, then \cite{Deloyeetal2007} note that it will 
delay the onset of donor's cooling and  increases the temperature of the donor.
So though this system lies below the  $T=0, M_\mathrm{a} = M_\mathrm{ch}$
 boundary, it does not necessarily mean that the donor can be approximated
as a zero-temperature object. }.

\begin{figure}[!hbp|t]
\centering
\includegraphics[height=18cm,width=13.0909cm,angle=270]{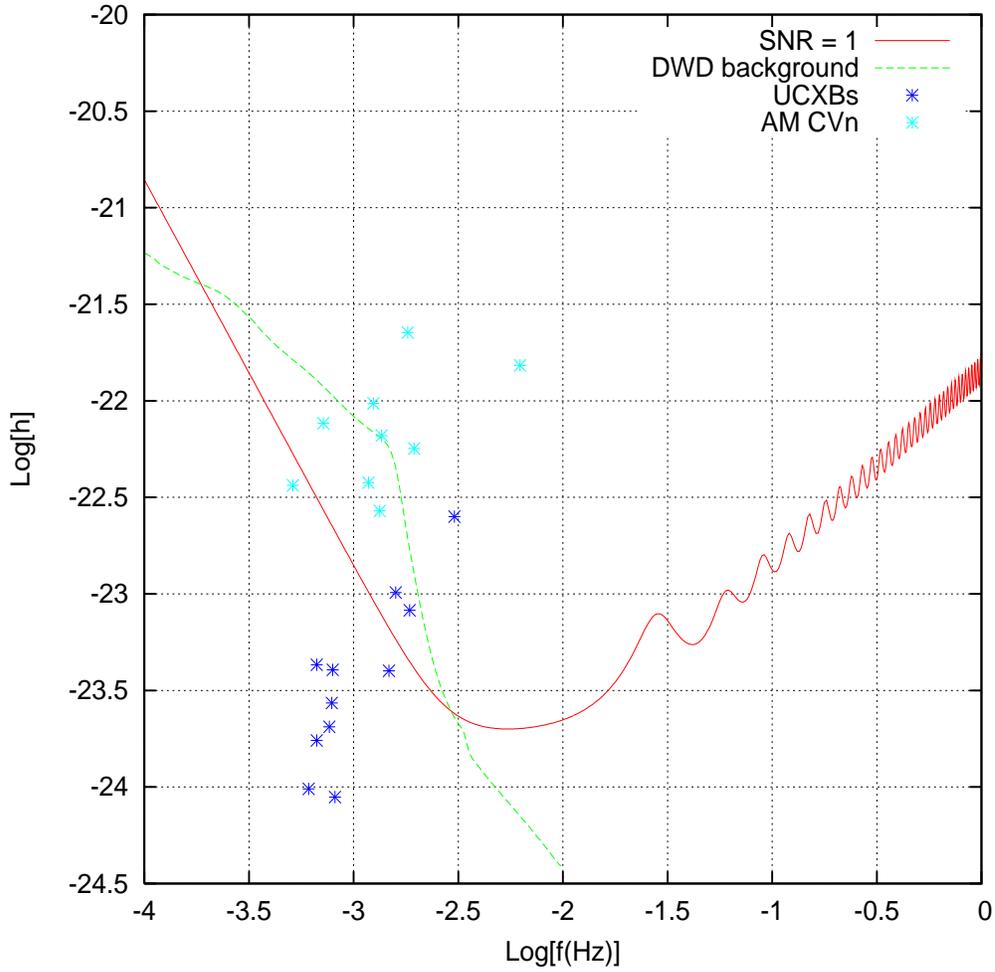}
\hfill \caption{ The known UCXBs (blue stars) and AM CVn systems (light blue
 stars) 
plotted on top of LISA's sensitivity curve for SNR = 1. In case of UCXBs, only one system
(4U 1820-30) has enough signal strength to be visible to LISA, whereas couple of known
AM CVns emit GWs above the instrumental and DWD background noise (green curve). }
\label{fig3}
\end{figure}

For NSWDs, similar to DWDs, the  contact boundaries  with He composition drawn  assuming 
 NS mass is $1.2 M_\odot$ (green curve with crosses ) and
 $3.0 M_\odot$ (green dashed curve)
 will also have a width due to decrease in $P_\mathrm{orb}$ even after the contact. Furthermore, 
the lower contact boundary lies {\it below} the upper contact boundary for DWDs
($M_\mathrm{a} = 1.44 M_\mathrm{\odot}$). This will make
the contact boundaries to ``overflow'' into the DWD region and LISA observations may not be able
to distinguish these two types of systems in this region. Here also the 
donor undergoes adiabatic evolution after $\dot M_\mathrm{d}$ reaches a maximum and consequently enters
the cooling phase at $P_\mathrm{orb}\approx 55$ min. Accordingly, in Fig. \ref{fig2}, the brown line shows
the orbital period of this transition phase. Beyond this line, the donors should start cooling and 
GW frequency slows down accordingly.
Fig. \ref{fig2} also shows the currently observed UCXBs, which we assume to be mass-transferring NSWD
binary systems. Table \ref{table1} gives the values of the masses of donors and orbital periods that
we used. Some of the UCXBs shown in Fig. \ref{fig2} lie {\it below} the lower contact boundary (green
cross curve), indicating that probably the donors are not of He composition. Although, this contact
boundary has a width, and hence they may have He composition, it is unlikely
 that we
 can observe them during this relatively short lived phase. Moreover, these systems are plotted assuming
the minimum mass of the donors mass range 
derived from observations, so this provides additional uncertainty in
determining the composition of donors. 
 It is unlikely that LISA will be able to observe  cooling donors in either DWD or NSWD systems
because of the instrumental and/or DWD background noise.
In Fig.\ref{fig3}, we plot the known UCXBs and AM CVns on top of LISA's
 sensitivity curve\footnote{http://www.srl.caltech.edu/~shane/sensitivity/MakeCurve.html}
 (SNR = 1) to assess the detectability of these systems. In the case of known UCXBs,
it is clear that only one system (4U 1820-30) has enough signal strength to be visible to LISA,
 whereas some of the known AM CVns emit GWs above the instrumental and DWD background noise.

Recalling from \S\ref{sec1},  in order to transform from log$(h_\mathrm{norm})$ 
 to log$(rh_\mathrm{norm})$  space,
 we need to know the distance $r$ to the binary system.
 The relation between the unknown binary parameters $r$, $M_\mathrm{tot}$
and $q$ and the observables $h_\mathrm{norm}$, $f$ and $\dot f$
 can be written as \citep{KT2007}
\begin{eqnarray}
\label{mass_r_relation}
\frac{M_\mathrm{tot}^{5}}{r^{3}}\biggl[\frac{q}{(1+q)^{2}}\biggr]^{3} &=& \frac{c^{12}}
{2^{6}\pi^{2}G^{5}}\frac{h_\mathrm{norm}^{3}}{f^{2}} , \\
r(1 - 2 g) &=& \frac{5 c}{24 \pi^{2}} \frac{\dot f}{h_\mathrm{norm} 
f^{3}}
\label{r_fdot_relation}
\end{eqnarray}
where $g=0$ in the inspiral phase of evolution and hence, it is easy to determine $r$
from  $h_\mathrm{norm}$, $f$ and $\dot f$ through Eq.(\ref{r_fdot_relation}).
 For mass-transferring systems $g$  is a function of $M_\mathrm{tot}$ and $q$, and they can
be related to the observable $f$ by the requirement that in the mass
 transfer phase, $R_\mathrm{d} = R_\mathrm{L}$. 


 The determination of $r$ and/or the masses of the stars in DWD/NSWD binary system 
depends on the determination of $\dot f$.  If an $\dot f$ can not be measured for a system,
 then there is no way to tell whether that system
is a NSWD or DWD system based only on LISA observations.
 If an $\dot f$ can be measured, and if it turns out to be negative, then it is possible 
that particular system is a mass-transferring system (DWD or NSWD). But as shown in Figs. \ref{fig1}
 $\&$ \ref{fig2}, it will still not
be possible to know the type of the system, at least for low frequency sources (Log$[f] \lessapprox -2$).
 There is a fairly good overlap in 
$\dot f$ between DWD $\&$ NSWD systems in this region because of the non-zero entropy nature
of the donors and also due to the lower limit on the mass of the NS ($1.2 M_\odot$).
But for high frequency mass-transferring sources (Log$[f] \gtrapprox -2$),
 it {\it may} still be possible to know the type of the system, as the overlap region 
reduces\footnote{There still will be some uncertainty for systems with NS mass lower than
 $M_\mathrm{ch}$, but for NS masses higher than  $M_\mathrm{ch}$,
 LISA should be able to distinguish both types of systems through the measurement of $\dot f$. }. 
 The same thing can be said about inspiralling systems because
    there is a large  overlap of
    DWD and NSWD inspirals at low frequencies  and even a measurement of positive
 $\dot f$ would unlikely be able distinguish these two types of inspiralling systems.

\section{Summary}
\label{sec4}
In the previous sections, we have outlined the construction of population boundaries
 to illustrate the various evolutionary phases
that a DWD or a NSWD binary system would undergo in the distance
 independent ``absolute'' amplitude-frequency domain (log$[rh_\mathrm{norm}]$ - log$[f]$) of LISA.
In an update to \cite{KT2007}, who assumed the donors to be fully degenerate $T=0$ He stars,  we 
consider that the donors in these systems  follow the M-R relationship of 
non-zero entropy donor models of \cite{DB2003} assuming He composition.
 These models assume fully convective and adiabatically
evolving donors during the whole episode of the mass-transfer phase. Figs. \ref{fig1} $\&$ \ref{fig2}
show that these
``hot'' donors occupy a larger fraction of the $rh - f$ space, than the $T=0$ donors, because hot 
donors are also more massive, increasing their intrinsic GW amplitude. At high frequencies, both the
models match each other because at these high masses, the hot donors are supported by degeneracy pressure
and the M-R relations match.
At low frequencies, the $T=0$ model and \cite{DB2003} $T=10^{4} K$ full model (which is equivalent to
$T=0$) diverge slightly because the Coulomb interactions dominate thermal contribution and M-R relations
deviate. 

 We also discussed the implications
of new evolutionary phases found by \cite{Deloyeetal2007} on our population boundaries. The initial
``turn on'' phase, where $\dot M_\mathrm{d}$ reaches it's maximum value and
 $P_\mathrm{orb}$ decreases even after contact, will result in
 contact boundaries having a width, instead of an abrupt cut-off.
 This will cause some overlap
 onto the C $\&$ O isotherms, which lie below He boundary. Soon after the system evolves from this 
initial contact, the second phase starts where the donor undergoes adiabatic evolution in response to the
mass-transfer and the system will follow a typical AM CVn evolutionary trajectory with decreasing GW 
amplitude and frequency. This will continue
until the donor begins to cool and the radius starts contracting. Accordingly, it  will reduce
 $\dot M_\mathrm{d}$,  and $P_\mathrm{orb}$ (and GW frequency) evolution slows down and may stall
 once the donor reaches fully degenerate configuration. But since LISA's
 sensitivity in this range is
limited by instrumental and DWD background noise, this part of the evolution (or cooling donors) will
probably be not observable by LISA.

It is unlikely that determination of $\dot f$ will shed light on the type
 (DWD/NSWD) of the low frequency systems 
without the help of independent electromagnetic observations.
This is because there is fairly a good overlap of NSWD and DWD systems within the resolvable frequency
regime (Log$[f] > 3$) of LISA \footnote{Since this region is occupied by both
 inspiralling $\&$ and mass-transferring systems, it may happen that more than one source can reside in
a frequency bin, but a measurement of $\dot f$ would probably indicate the evolutionary phase.}.  But for high frequency sources, it might be possible to 
distinguish between them.
Combined with the expectation
 that the relative population of NSWD is low compared to DWDs and short period systems
do not stay longer at those periods, it is likely that
LISA will be able to measure $\dot f$ for more number of DWDs than NSWDs. 
 It will be
interesting to see how the high frequency regions constrained by these
boundaries would be populated through LISA observations and whether indeed
we will know the nature of these systems.

$\\$

\acknowledgements

We thank an anonymous referee whose suggestions have led to significant
 improvements in the
manuscript. R. K gratefully acknowledges the support of National Science
 Foundation Grant
No.~PHY 06-53462 and No.~PHY 05-55615, and NASA Grant No.~NNG05GF71G,
awarded to The Pennsylvania State University.
Many thanks to Joel Tohline (LSU) for the help and guidance provided in the 
preparation of this work and manuscript.

\begin{deluxetable}{cccccc}
\tablecaption{Observed and derived properties of some of the known UCXBs \citep{intzand2007} shown
 in Fig.\ref{fig2}. $M_\mathrm{d}$ indicates minimum mass of the donor. 
References: (1) \cite{Tarana2007} (1A) \cite{Cumming2003} (2) \cite{WC2004} 
(2A) \cite{WC2004} (3) \cite{Sidoli2006} (3A) \cite{JC2005} (4) \cite{Dieball2005} (4A) \cite{McNamara2004} (5) \cite{Falanga2005} (5A) \cite{Campana2003} (6) \cite{Krauss2007} (6A) \cite{C1998} (7)\cite{Gierlinski2005} (7A) \cite{Papitto2008} (8) \cite{Galloway2002} (8A) \cite{Galloway2002} (10A)  \cite{C1998} 
(11A) \cite{Nelemans2004} (12) \cite{Krimm2007} (12A) http://web.mit.edu/newsoffice/2007/pulsar-0913.html}
\tablewidth{0pt}
\tablehead{ \colhead{Name} & \colhead{Orbital period} & \colhead{$\frac{M_\mathrm{d}}{M_\odot}$}& \colhead{Reference} & \colhead{distance} &\colhead{Reference} \\
& \colhead{(min)} & & \colhead{for mass} & \colhead{(kpc)} & \colhead{for distance}}
\startdata 4U 1820-30 & 11 & 0.06 & 1 & 7.6 & 1A\\
4U 1543-624 & 18 & 0.025 & 2 & 7.0 & 2A\\
4U 1850-087 & 21 & 0.04 & 3 & 8.2 & 3A\\
M15 X-2 & 22.6 & 0.02 & 4 & 9.98 & 4A\\
XTE J1807-294 & 41 & 0.0053 & 5 & 8.0 & 5A\\
4U 1626-67 & 42 & 0.04 & 6 & 5 & 6A\\
XTE J1751-305 & 42.4 & 0.014 & 7 & 6.7 & 7A\\
XTE J0929-314 & 43.6 & 0.008 & 8 & 5.0 & 8A\\
NGC 6652B & 43.6 & ? & -- & ? & --\\
4U 1916-05 & 50 & ? & -- & 8.9 & 10A\\
4U 0614+091 & 50 & ? & -- & 3.0 & 11A\\
SWIFT J1756.9-2508 & 54.7 & 0.0067 & 12 & 7.6 & 12A\\
 \\
\enddata
\tablecaption{this is}
\label{table1}
\end{deluxetable}


\clearpage
\thispagestyle{empty}

\end{document}